%% file: 0paper.tex
\documentclass[format=sigconf, screen=true]{acmart}

\renewcommand\footnotetextcopyrightpermission[1]{} 
\input{0packagesNew.tex}

\AtBeginDocument{%
  \providecommand\BibTeX{{%
    \normalfont B\kern-0.5em{\scshape i\kern-0.25em b}\kern-0.8em\TeX}}}

\setcopyright{none}

\begin{document}






\title[AI Chatbots for Mental Health: Values and Harms from Lived Experiences of Depression]{AI Chatbots for Mental Health: Values and Harms from\\ Lived Experiences of Depression}





\author{Dong Whi Yoo}
\orcid{0000-0003-2738-1096}
\affiliation{%
 \institution{Indiana University Indianapolis}
 \city{Indianapolis}
 \state{IN}
 \country{USA}}
 \email{dongwhi.yoo@gmail.com}
 
\author{Jiayue Melissa Shi}
\orcid{0009-0007-0624-2421}
\affiliation{%
 \institution{University of Illinois Urbana-Champaign}
 \city{Urbana}
 \state{IL}
 \country{USA}}
\email{mshi24@illinois.edu}

\author{Violeta J. Rodriguez}
\orcid{0000-0001-8543-2061}
\affiliation{%
 \institution{University of Illinois Urbana-Champaign}
 \city{Urbana}
 \state{IL}
 \country{USA}}
\email{vjrodrig@illinois.edu}

\author{Koustuv Saha}
\orcid{0000-0002-8872-2934}
\affiliation{%
 \institution{University of Illinois Urbana-Champaign}
 \city{Urbana}
 \state{IL}
 \country{USA}}
\email{ksaha2@illinois.edu}

\renewcommand{\shortauthors}{Yoo et al.}


\input{0abstract}



\maketitle


\input{1introduction.tex} 
\input{2rw.tex} 
\input{3method} 
\input{4findings} 

\input{5discussion} 
\input{6conclusion} 




\bibliographystyle{ACM-Reference-Format}
\bibliography{0paper}



\end{document}

\endinput

%% file: 0packagesNew.tex
\usepackage{color, colortbl, xcolor}
\usepackage{url}
\usepackage{subcaption}
\usepackage{textcomp}
\usepackage{soul}
\usepackage{multirow}
\usepackage{enumitem}
\usepackage{mathtools}
\usepackage{siunitx}

\usepackage{graphicx} 
\usepackage{placeins} 

\usepackage{subcaption}
\usepackage{graphicx}
\usepackage{lipsum}
\usepackage{caption}

\usepackage{array} 
\usepackage{ragged2e} 

\usepackage{tikz}
\usepackage{xcolor}

\usepackage{booktabs} 
\usepackage{graphicx} 
\usepackage{graphicx}  

\usepackage{arydshln}
\definecolor{linkColor}{RGB}{6,125,233}
\definecolor{green}{rgb}{0.0, 0.65, 0.31}
\definecolor{bleudefrance}{rgb}{0.19, 0.55, 0.91}
\definecolor{ceruleanblue}{rgb}{0.16, 0.32, 0.75}
\definecolor{grey}{HTML}{969696}
\definecolor{violet}{HTML}{756bb1}
\definecolor{dgrey}{HTML}{525252}
\definecolor{lgrey}{HTML}{5ab4ac}
\definecolor{dgreen}{HTML}{005a32}
\definecolor{purple}{HTML}{ae017e}

\definecolor{lethalcolor}{HTML}{e66101}
\definecolor{nlethalcolor}{HTML}{5e3c99}
\definecolor{LightGray}{gray}{0.97}

\definecolor{editCol}{HTML}{000000}
\definecolor{maskCol}{HTML}{c51b7d}
\definecolor{lrColor}{HTML}{8856a7}
\definecolor{trColor}{HTML}{d01c8b}
\definecolor{ctColor}{HTML}{4dac26}
\definecolor{brickred}{HTML}{f03b20}
\definecolor{improveCol}{HTML}{253494}
\definecolor{worsenCol}{HTML}{d7191c}
\definecolor{DarkBlue}{HTML}{00008B}
\definecolor{mscolor}{HTML}{01665e}
\definecolor{nmscolor}{HTML}{bf812d}
\definecolor{lgreen}{HTML}{ccece6}
\definecolor{dolive}{HTML}{308014}
\definecolor{lgreen}{HTML}{e0f3db}
\definecolor{dpink}{HTML}{CD1076}
\definecolor{pink}{HTML}{FED2D2}
\definecolor{soothinggreen}{HTML}{4dac26}
\definecolor{darkred}{HTML}{8B0000}

\colorlet{tablerowcolor4}{gray!50} 

\newcommand*{\textlabel}[2]{%
  \edef\@currentlabel{#1}
  \phantomsection
  #1\label{#2}
}

\colorlet{tableheadcolor}{gray!25} 
\colorlet{tablerowcolor}{gray!10} 
\colorlet{tablerowcolor2}{gray!45} 
\colorlet{tablerowcolor3}{gray!25} 

\newcommand{\rowcollight}{\rowcolor{tablerowcolor3}} %

\newcolumntype{a}{>{\columncolor{tablerowcolor}}r}
\definecolor{aicolor}{HTML}{018571}
\definecolor{occolor}{HTML}{ff7799}

\definecolor{aicolor}{HTML}{fc8d62}
\definecolor{occolor}{HTML}{253494}

\newif{\ifhidecomments}
  \hidecommentsfalse 
\ifhidecomments
    \newcommand{\dongwhi}[1]{}
    \newcommand{\koustuv}[1]{}
\else
    \newcommand{\dongwhi}[1]{\textbf{\small\sffamily{\textcolor{marroon}{[#1 -- Dong Whi]}}}}
    \newcommand{\koustuv}[1]{\textbf{\small\sffamily{\textcolor{dpink}{[#1 -- Koustuv]}}}}
  \fi
\newcommand{\edit}[1]{{\textcolor{editCol}{#1}}}

\newcommand{\ai}{AI}

\renewcommand{\textrightarrow}{$\rightarrow$}

\newcommand{\tabitem}{\textbullet~~}

\colorlet{tableheadcolor}{gray!25} 

\definecolor{neutralCol}{HTML}{dd1c77}
\definecolor{neutralGreen}{HTML}{31a354}
\definecolor{NewBlue}{HTML}{1879ba}
\definecolor{bleudefrance}{rgb}{0.19, 0.55, 0.91}  
\definecolor{AfTrColor}{HTML}{0868ac}  
\definecolor{BfTrColor}{HTML}{a8ddb5}  

\definecolor{AfCtColor}{HTML}{b10026}  
\definecolor{BfCtColor}{HTML}{fd8d3c}

\graphicspath{ {figures/} }



%% file: 0abstract.tex
\begin{abstract}

Recent advancements in LLMs enable chatbots to interact with individuals on a range of queries, including sensitive mental health contexts. Despite uncertainties about their effectiveness and reliability, the development of LLMs in these areas is growing, potentially leading to harms. To better identify and mitigate these harms, it is critical to understand how the values of people with lived experiences relate to the harms. In this study, we developed a technology probe---a GPT-4o based chatbot called Zenny---enabling participants to engage with depression self-management scenarios informed by previous research. We used Zenny to interview 17 individuals with lived experiences of depression. Our thematic analysis revealed key values: informational support, emotional support, personalization, privacy, and crisis management. This work explores the relationship between lived experience values, potential harms, and design recommendations for mental health AI chatbots, aiming to enhance self-management support while minimizing risks.

\end{abstract}

%% file: 1introduction.tex
\section{Introduction}\label{section:intro}

The recent surge in Large Language Models (LLMs) and AI-powered chatbots such as OpenAI's ChatGPT, Google's Gemini, and Microsoft's Bing Search has sparked widespread interest in exploring their application across various fields, including healthcare~\cite{jo2024understanding,sharma2024facilitating,saha2025ai}. 
In mental health, LLMs are expected to address issues such as resource scarcity, subjective diagnoses, and stigmatization~\cite{lai2023psyllm}. 
One of the most popular mental health chatbots, Woebot, is incorporating LLMs into its system, expecting to enhance its existing rule-based models~\cite{fitzpatrick2017delivering, woebot_ai}.
Furthermore, empirical evidence shows that 
people with mental health concerns are using AI-based chatbots for mental health purposes~\cite{ma2024evaluating, song2024typing}.

Despite these technological advances and the growing use of AI in mental health, ethical concerns remain significant. \citeauthor{cabrera2023ethical} highlighted moral dilemmas through bioethical principles~\cite{cabrera2023ethical}, and \citeauthor{de2023benefits} highlighted the potential benefits and harms within an ecological framework~\cite{de2023benefits}. 
Accordingly, it is important to enhance our understanding of potential harms by incorporating the values and voices of the most affected individuals---those with lived experiences of mental health challenges. This approach can inform designs that recognize the unique, personal aspects of mental health and ensure technologies minimize potential risks~\cite{de2023benefits}.

In the field of human-computer interaction (HCI), researchers have extensively explored the values users prioritize and the potential harms technologies might bring~\cite{foong2024designing, sadek2024guidelines, friedman2013value, boyd2022designing, yoo2017stakeholder,xu2023transitioning}. Values are defined as ``what a person or group considers important in life'' \cite{friedman2013value}, encompassing both tangible elements, such as friends, and more abstract concepts, such as virtue. Understanding these values is essential for anticipating and mitigating the potential harm that new technologies might introduce. 
In particular, Value-Sensitive Design (VSD) offers systematic methods for incorporating user values into technology design~\cite{friedman1996value}. Initially, VSD emphasized a theoretical understanding of universal and important user values through triangulated approaches that included theoretical, empirical, and technical studies~\cite{friedman1996value}. Later, \citeauthor{le2009values} argued that user values are situated and contextual, suggesting that researchers should prioritize understanding the lived experiences of users to better support their values and mitigate potential harms~\cite{le2009values}. 
This motivates us to adopt this perspective of values as lived experiences to identify potential harms in the context of LLMs in mental health. 

For the scope of this study, we focus on the self-management of depression. Depression, or major depressive disorder, is one of the most prevalent mental health conditions, affecting approximately 8\% of U.S. adults~\cite{MHA_Depression}. It is also one of the most common comorbid conditions.
We specifically focus on self-management contexts because individuals with lived experiences of depression must manage their symptoms daily, regardless of the severity of their condition or treatment process~\cite{van2015patients}. 
Therefore, we target the following research questions (RQs):

\begin{enumerate}
    \item[\textbf{RQ1:}] What values do people with depression prioritize when using AI chatbots to self-manage depression?
     \item[\textbf{RQ2:}] What potential harms might AI chatbots present in relation to the values identified above?
\end{enumerate}

We first built a technology probe~\cite{hutchinson2003technology} where our participants could interact with Zenny---a chatbot based on GPT-4o---under hypothetical scenarios related to depression self-management. 
We designed these scenarios based on prior work that defined self-management strategies from the perspectives of lived experiences~\cite{van2015patients}.
We recruited 17 participants who self-reported lived experiences of depression. They engaged with Zenny across four different self-management scenarios~\cite{van2015patients}. We analyzed the chat history and responses to follow-up questions, to identify five key values: 1) informational support, 2) emotional support, 3) personalization, 4) privacy, and 5) crisis management. 
Reflecting on these findings, we articulate potential harms and offer design recommendations for future AI chatbots for mental health. Additionally, we propose a harm mitigation checklist to help practitioners, designers and researchers implement these recommendations effectively. 
Therefore, our work contributes \textbf{a mapping of the potential harms of LLMs to the values of people with lived experiences, and offers design recommendations for future mental health chatbots}. 
This is especially important because we found a stark difference between the excitement expressed by our participants and the concerns raised in prior work~\cite{cabrera2023ethical, de2023benefits} regarding the use of AI chatbots for mental health. 
Despite identifying the significant harms, public and industry enthusiasm may drive widespread proliferation and adoption of AI chatbots, including in sensitive cases like healthcare.
As researchers, it is our responsibility to anticipate potential harms and raise awareness among various stakeholders to prevent and mitigate the negative consequences.

%% file: 2rw.tex
\section{Background and Related Work}\label{section:rw}

\subsection{AI Ethics and Values: Anticipating Harms}



Autonomous technologies, such as robots and AI agents, raise ethical and value-based questions~\cite{winner1978autonomous} because they may perform tasks and exhibit behaviors that were not explicitly programmed. The values these technologies operate under are critical in determining whether they are beneficial or harmful when carrying out these less-regulated tasks. HCI researchers have explored the values and ethics of these technologies~\cite{cheon2016integrating,cheon2018futuristic}. 
A logical subsequent question concerns how the values of autonomous technologies align with those of the individuals they impact.
As AI continues to influence areas such as employment~\cite{kawakami2023wellbeing,dasswain2023algorithmic,roemmich2023emotion,das2024teacher}, immigration~\cite{rohanifar2022kabootar,chen2020creating}, social welfare~\cite{kawakami2022care}, and political debate~\cite{andric2023reconciling}, research highlights that when the values embedded in these technologies favor the powerful (e.g., employers) over those directly affected (e.g., employees), significant harm may ensue.

Relatedly, \citeauthor{muller2017exploring} emphasized the importance of understanding the values of those who will be impacted by AI technologies developed by privileged developers and researchers:
    ``The future users of these technologies will inhabit a third world of AI, in which ordinary people must deal with AI entities that are produced by other people, and inevitably reflect the interests of those other people. How can users begin to write their own accounts of technologies they envision, and values that are implicated by those future technologies?''~\cite{muller2017exploring}


Value-Sensitive Design (VSD)~\cite{friedman1996value} has provided a systematic approach to understanding stakeholder values and guiding the design of technologies that reflect and support these values. \citeauthor{friedman1996value} proposed a triangulated approach that integrates theoretical, empirical, and technical perspectives to understand stakeholder values~\cite{friedman1996value}. Later, \citeauthor{le2009values} emphasized that stakeholder values are situated and contextual, highlighting the importance of working with individuals with lived experiences~\cite{le2009values}. 
Our study builds on value-based approaches, recognizing that AI chatbots for mental health must reflect and uphold the values of people with lived experiences.

Building on the principles of VSD, \citeauthor{zhu2018value} proposed Value-Sensitive Algorithm Design (VSAD)~\cite{zhu2018value}, which emphasizes the importance of engaging stakeholders in the early stages of algorithm development. 
This approach aims to avoid biases in design choices and ensure that stakeholders' values are not compromised~\cite{zhu2018value}. 
By involving stakeholders from the outset, VSAD seeks to create algorithms that are more equitable and aligned with the diverse needs of those affected by them. \citeauthor{madaio2020co} further advanced this idea by adopting a participatory approach to co-designing ethical AI checklists~\cite{madaio2020co}. 
These checklists were developed in collaboration with stakeholders to address organizational challenges and promote a deeper understanding of AI ethics~\cite{madaio2020co}. 
This participatory approach not only helps to surface hidden biases but also fosters a sense of ownership and accountability among those involved in the design process---as also highlighted in a body of research in this space~\cite{ehsan2023charting,kawakami2023wellbeing,roemmich2023emotion,yoo2024missed}.

Another important discussion related to AI and human values is AI alignment, which refers to ensuring that AI systems produce desirable outcomes without unintended negative consequences~\cite{terry2023ai, goyal2024designing}. 
Originally stemming from philosophical approaches to AI systems~\cite{gabriel2020artificial}, AI alignment has been widely discussed in machine learning and AI research and is now emerging in HCI literature~\cite{shen2024towards, terry2023ai,suh2024luminate}. 
AI alignment often addresses how AI systems align with human values during the development process (e.g., annotation) or through interaction (e.g., human feedback)~\cite{li2023coannotating,chung2022talebrush, hota2024cognitively}. \citeauthor{shen2024towards} introduced the concept of bidirectional alignment, which includes HCI research efforts in AI education, critical thinking about AI, and human-AI collaboration~\cite{shen2024towards}. 

\citeauthor{khoo2024s} revealed that users' privacy concerns often differ from those of experts. For instance, experts tend to focus on technical aspects like encryption and password strength, whereas users emphasize practical concerns such as HIPAA violations and exploitative practices~\cite{khoo2024s}. 
Similarly, although the potential harms of LLMs in mental health have been investigated, these discussions are predominantly top-down and driven by expert opinions. 
In contrast, we empirically consolidate the perspectives of individuals with lived experiences of depression by understanding their values and mapping potential harms.
By mapping the values of individuals with lived experiences \edit{of depression} to potential harms and providing a harm mitigation checklist, we contribute to the existing literature on ethical AI guidelines~\cite{amershi2019guidelines,liao2020questioning,kaur2022sensible,madaio2020co,mamykina2022grand,dhanorkar2021needs} and value-sensitive AI~\cite{muller2017exploring,zhu2018value,ballard2019judgment}, offering a specific application case in mental health and a deeper understanding of how VSD can support the values of those with lived experiences \edit{of depression}.

\subsection{Chatbots and LLMs for Mental Health}
\edit{Chatbots, with their conversational affordances, are expected to address challenges in mental health, such as the lack of sufficient resources~\cite{sweeney2021can,vaidyam2019chatbots}. Before widespread adoption of LLMs, mental health chatbots primarily relied on rule-based systems and employed various therapeutic approaches to guide users through self-directed exercises~\cite{sweeney2021can, lee2020designing, dubiel2022conversational, collins2022covid, park2021wrote,kang2024app}. Notably, \citeauthor{wester2024chatbot} demonstrated that perceived moral agency, as evidenced in multiple-choice chatbots, significantly impacts trust, likeability, and safety~\cite{wester2024chatbot}. Beyond moral agency, keyboard typing patterns have been utilized to detect emotions and support text-based mental health services~\cite{sim2024towards}. Also, ChatPal, a multiple-choice chatbot designed for individuals in rural areas, provided valuable insights into usability and user experience~\cite{boyd2022usability}. Further, \citeauthor{ng2023emotion} showed that incorporating cultural adaptation and emotion detection can enhance the effectiveness of non-LLM-based mental health chatbots~\cite{ng2023emotion}.}

Recent advances in generative AI and LLMs have led to increased interest in their applications for several healthcare contexts, including mental health support~\cite{xu2021chatbot,goodman2022lampost,tseng2023understanding,mitchell2022examining,lai2023towards,jo2023understanding,saha2025ai}. 
LLM-based chatbots could facilitate open dialogues that parse user semantics, allowing for emotionally resonant interaction~\cite{ma2023understanding}. 
This capability is particularly beneficial for individuals who may lack access to traditional mental health resources due to geographical, temporal, or social constraints. 
Additionally, studies have shown that LLM-based conversational agents can support mental health by offering on-demand, non-judgmental assistance, fostering social confidence, and encouraging self-discovery~\cite{sweeney2021can,sabour2023chatbot,crasto2021carebot,o2023human,haque2023overview,rao2024integrating,ma2023understanding,lee2020hear}.
Prior work designed and evaluated the efficacy of chatbots that encourage self-disclosure and expressive writing~\cite{park2021wrote,lee2020designing,lee2020hear}.


AI chatbots can infer the contexts of input texts (i.e. prompts) and generate responses that align coherently with these prompts. People use LLMs' abilities to build information extraction~\cite{pilault2020extractive}, and code generation systems~\cite{chen2021evaluating}. Following these abilities, LLMs are increasingly being integrated into systems that provide mental health support, engaging users in open dialogues that parse the semantics of their inputs and interact with them on an emotional level~\cite{de2023benefits, crasto2021carebot, o2023human, choi2024unlock,haque2023overview}. The use of LLMs in chatbots, with their clean and user-friendly conversational interfaces, has generated considerable excitement among clinicians about the potential for innovative AI methods of delivering mental health support. 

Despite these advances, significant concerns remain about using LLMs for mental health, including ethical issues like information accuracy, risks of emotional distress, and the potential for user over-reliance on technology~\cite{de2023benefits,cabrera2023ethical}. \citeauthor{cabrera2023ethical} called for interdisciplinary collaboration to establish effective guidelines and regulatory frameworks that ensure the responsible use of chatbots while safeguarding user well-being and promoting ethical standards in mental health care~\cite{cabrera2023ethical}. While not explicitly aimed at preventing harms, \citeauthor{weisz2024design} developed design principles for generative AI applications that can be understood as contributing to harm prevention~\cite{weisz2024design}. These principles include designing responsibly and fostering appropriate trust and reliance~\cite{ma2023should}.
Building upon this body of work, our research aims to 1) anticipate the potential harms associated with using LLM-based chatbots in the context of mental health, and 2) map these potential harms with the values of people with lived experiences of mental health conditions. 
This work is particularly critical as interactive LLMs (e.g., ChatGPT, Gemini in Google Search, Bing Search, etc) are becoming increasingly pervasive and widely used for everyday information-seeking purposes. Without a comprehensive understanding of how sensitive populations perceive and engage with this technology, preventing and mitigating the potential harms and downstream effects will remain a significant challenge.

\subsection{Depression Research in HCI}
Depression is a widespread mental health issue impacting millions of people around the world. The World Health Organization (WHO) reports that over 264M people worldwide experience depression~\cite{who2023mental,nepal2024moodcapture}. 
Depression can severely impact emotional wellbeing~\cite{rottenberg2005emotion}, physical health~\cite{frerichs1982physical} and social relationships~\cite{santini2015association}. In more extreme cases, depression can lead to suicide, with approximately 800,000 deaths occurring each year~\cite{amaltinga2020factors, bertolote2003suicide}. Additionally, the 2020 National Survey on Drug Use and Health reported that approximately 1 in 10 Americans and 1 in 5 young adults and adolescents disclosed experiencing depression~\cite{felgenhauer2021same}. Research also suggests that the incidence of depression nearly tripled during the COVID-19 pandemic~\cite{lamberton2016new}, with projections estimating that the rising prevalence of mental health conditions will cost the global economy around \$1 trillion~\cite{lederman2019support}. 

When managing depression, self-management strategies are crucial because individuals with depression often need to cope with their symptoms outside of clinical settings (e.g., between clinical encounters), and the outcomes of depression management are closely linked to everyday behaviors (e.g., going to work, cleaning the house)~\cite{spors2021selling}. \citeauthor{houle2013depression} conducted a systematic review exploring various self-management approaches for depression and assessing their efficacy, finding that self-management is associated with reduced depressive symptoms and improved functioning~\cite{houle2013depression}. Additionally, the American Psychiatric Association's depression guidelines briefly mention self-management in the context of incomplete recovery: ``the professional should add a disease management component to the overall treatment plan [...] such as developing self-management skills~\cite{american2018american}.'' However, many of the existing efforts to define and understand self-management strategies are clinician- or researcher-driven, leaving out the voices and perspectives of those who actually practice these strategies. To address this gap, \citeauthor{van2015patients} engaged individuals with depression to define and conceptualize self-management strategies, resulting in 50 strategies reported to be effective by people with lived experience\cite{van2015patients, van2018use}. This bottom-up approach to understanding self-management strategies laid the foundation for our study, and we built our hypothetical scenarios based on the most relevant strategies from their work~\cite{van2015patients}.

In the context of HCI, 
prior work studied the design and usability of digital interventions in impacting individuals suffering from depression disorders~\cite{sas2020mental}. 
HCI researchers have actively worked to support people with lived experiences of depression~\cite{yoo2023discussing, sas2020manneqkit}. \citeauthor{nepal2024moodcapture} developed machine learning models with in-the-wild smartphone images to identify depressive symptoms~\cite{nepal2024moodcapture}. Additionally, \citeauthor{wu2024designing} investigated the design and integration of online peer support groups (PSGs) as adjuncts to evidence-based interventions for parents of adolescents facing mental health challenges. 
\citeauthor{bhattacharjee2023investigating} highlighted the importance of understanding and incorporating context when delivering mental health interventions~\cite{bhattacharjee2023investigating}. \citeauthor{sogaard2019combining} underscored the importance of interdisciplinary collaborations between mental health professionals and computing scientists to create tailored interventions that address the specific needs of patients with depression. 
Our research focuses on exploring people's opinions on using AI chatbots for mental health and depression self-management. 
We designed our study to examine user perspectives about a near-realistic use case of interactions with a mental health chatbot. 
We integrated theory-driven scenario-based interactions~\cite{van2015patients} as well as an open-ended ``Own question'' option.
Our study aims to contribute to the understanding of how people with lived experiences interact with AI tools for self-managing their mental health.

%% file: 3method.tex
\section{Interview Study using Technology Probe}

Our study design consists of a scenario-based interview study with a technology probe~\cite{hutchinson2003technology} (a chatbot named Zenny) with participants who have had lived experiences with depression.
This study was approved by the respective Institutional Review Boards (IRBs) at the researchers' institutions.
In the following subsections, we describe our methodology. 


\subsection{Zenny: Technology Probe for Depression Self-Management}
One of the challenges in understanding user values of emerging technologies is that many people lack sufficient experience with these technologies~\cite{odom2012fieldwork}. To explore this design space that they may not have lived in~\cite{muller2017exploring}, we built a technology probe~\cite{hutchinson2003technology} in our study. The primary purpose of a technology probe is to collect user data in real-world settings to better understand how people use technologies and how future technologies can better support them~\cite{hutchinson2003technology}. However, due to the sensitive and potentially risky nature of depression self-management, we chose not to deploy our technology probe named Zenny---a GPT-4-based chatbot---in a real-world setting. Instead, we asked our participants to interact with Zenny during a one-hour interview session. During these sessions, they were presented with four different depression self-management strategies derived from previous literature~\cite{van2018use, van2015patients}. By interacting with Zenny within specific scenarios, participants were able to share their experiences and opinions in a controlled, less stimulating environment. In the subsections, we elaborate on the scenario design process and the development of the chatbot.

\subsubsection{Scenario Design}

\begin{table*}[t!]
    \centering
    \sffamily
    \small
        \caption{The four scenarios for the interviews were based on literature regarding the lived experience of depression self-management~\cite{van2015patients}. Participants completed two to four scenarios, depending on their speed of interaction and the available time during the interviews. The order of the four scenarios was randomly assigned. Participants read aloud the descriptions and tasks, asking the interviewer questions if they had any. They were allowed to interact back and forth with Zenny.}
    \label{tab:zenny_scenarios}
    \begin{tabular}{p{0.3\columnwidth} p{0.85\columnwidth} p{0.83\columnwidth}}
    \textbf{Scenario} & \textbf{Description} & \textbf{Task}\\
    \toprule
    1. Goal setting & You have been feeling overwhelmed and discouraged lately due to depression. You want to take steps to improve your daily life and know that setting small, achievable goals can be a helpful way to start. But, you are unsure where to begin or what kind of goals would be both realistic and supportive. & Use Zenny to help you brainstorm a list of 3-5 realistic, short-term goals that can boost your mood and help you manage your day-to-day life while living with depression. As you interact with Zenny, consider discussing strategies for setting achievable goals, tracking progress, and making adjustments as needed.\\
    \rowcolor{LightGray} \raggedright 2. Discussing depression with those you trust & You have been living with depression. You want to explain what you are going through to someone you trust – a friend, family member, or partner. You hope this conversation will help them understand you better and give you the support you need, but you are not sure how to start. &  Use Zenny to get advice on how to have a conversation about your depression with someone you trust. Focus on getting tips for starting the conversation, explaining your experience, and asking for support. \\
   \raggedright  3. Leaving the house regularly & Depression has been making you want to stay inside. You know getting out more could help, but it feels hard. Planning activities, or even just going outside, can seem overwhelming. Yet, you want to find ways to make small outings part of your life. & Talk to Zenny about how to start getting out of the house more often to help manage your depression, focusing on easy activities, motivation, planning, setting expectations, and tracking progress. \\
    \rowcolor{LightGray} \raggedright 4. Finding a different therapist when there is limited progress & You have been in therapy for a while. It helped at first, but lately, you feel stuck. The things that used to work do not seem to help anymore. You are thinking about finding a new therapist who might have a fresh perspective. But, starting over feels overwhelming, and you are not sure how to find the right person. & Use Zenny to get advice on how to know when it is time to find a new therapist, what to look for in a new therapist, and how to navigate the process of switching therapists. \\
    \bottomrule
    \end{tabular}

    \Description[table]{The table described four different scenarios used in interviews. Each scenario was detailed with a description and an associated task that involved interacting with our AI-based chatbot Zenny. These scenarios were driven by the literature on the lived experience of depression self-management. Scenario 1: Goal Setting. Description: You've been feeling overwhelmed and discouraged lately due to depression. You want to take steps to improve your daily life and know that setting small, achievable goals can be a helpful way to start. But, you're unsure where to begin or what kind of goals would be both realistic and supportive. Task: Use ChatGPT to help you brainstorm a list of 3-5 realistic, short-term goals that can boost your mood and help you manage your day-to-day life while living with depression. As you interact with ChatGPT, consider discussing strategies for setting achievable goals, tracking progress, and making adjustments as needed.
Scenario 2: Discussing depression with those you trust. Description: You've been living with depression. You want to explain what you're going through to someone you trust – a friend, family member, or partner. You hope this conversation will help them understand you better and give you the support you need, but you're not sure how to start.
Task: Use ChatGPT to get advice on how to have a conversation about your depression with someone you trust. Focus on getting tips for starting the conversation, explaining your experience, and asking for support. Scenario 3: Leaving the house regularly. Description: Depression has been making you want to stay inside. You know getting out more could help, but it feels hard. Planning activities, or even just going outside, can seem overwhelming. Yet, you want to find ways to make small outings part of your life. Task: Talk to ChatGPT about how to start getting out of the house more often to help manage your depression, focusing on easy activities, motivation, planning, setting expectations, and tracking progress. Scenario 4: Finding a different therapist when there is limited progress. Description: You've been in therapy for a while. It helped at first, but lately, you feel stuck. The things that used to work don't seem to help anymore. You're thinking about finding a new therapist who might have a fresh perspective. But, starting over feels overwhelming, and you're not sure how to find the right person. Task: Use ChatGPT to get advice on how to know when it's time to find a new therapist, what to look for in a new therapist, and how to navigate the process of switching therapists.}

\end{table*}


We designed the scenarios to guide our participants in interacting with an AI chatbot for depression self-management. 
We prepared these scenarios to avoid asking participants to query the chatbot directly based on their lived experiences, which could be potentially sensitive, unpredictable, and triggering. 
Instead, we prompted participants to reflect on theory-driven
self-management strategies within the limited interview time.

\citeauthor{van2015patients} conducted focus groups with people with lived experiences of depression to identify their self-management strategies; finding 50 of the most common strategies grouped into eight different themes~\cite{van2015patients}.
From these themes, we identified the four most relevant to AI chatbots---
1) \textit{daily life strategies and rules}, 2) \textit{explanation of the disease to others}, 3) \textit{engaging in activities}, and 4) \textit{proactive attitude towards depression and treatment}. 
Then, we selected one strategy from each of these themes to develop four scenarios: 
1) goal setting, 2) discussing depression with trusted ones, 
3) leaving the house regularly, and 4) finding a different therapist when there is limited progress.




We developed detailed descriptions for each scenario to help participants understand the context, along with specific tasks that encouraged interaction with the chatbot. These scenarios and their descriptions are presented in \autoref{tab:zenny_scenarios}. During the interviews, the scenarios were presented to participants in random order, and they were asked whether they found each scenario relevant to their own experience. If a participant considered the scenario relevant or had something they wanted to ask, they interacted with the Zenny; otherwise, they skipped that specific scenario. The number of scenarios presented to each participant varied depending on the pace of their interaction with Zenny. In the next section, we elaborate on the development of the Zenny.

\begin{figure*}[t]
    \centering
    \begin{subfigure}[t]{0.98\columnwidth}
        \centering
        \includegraphics[width=\columnwidth]{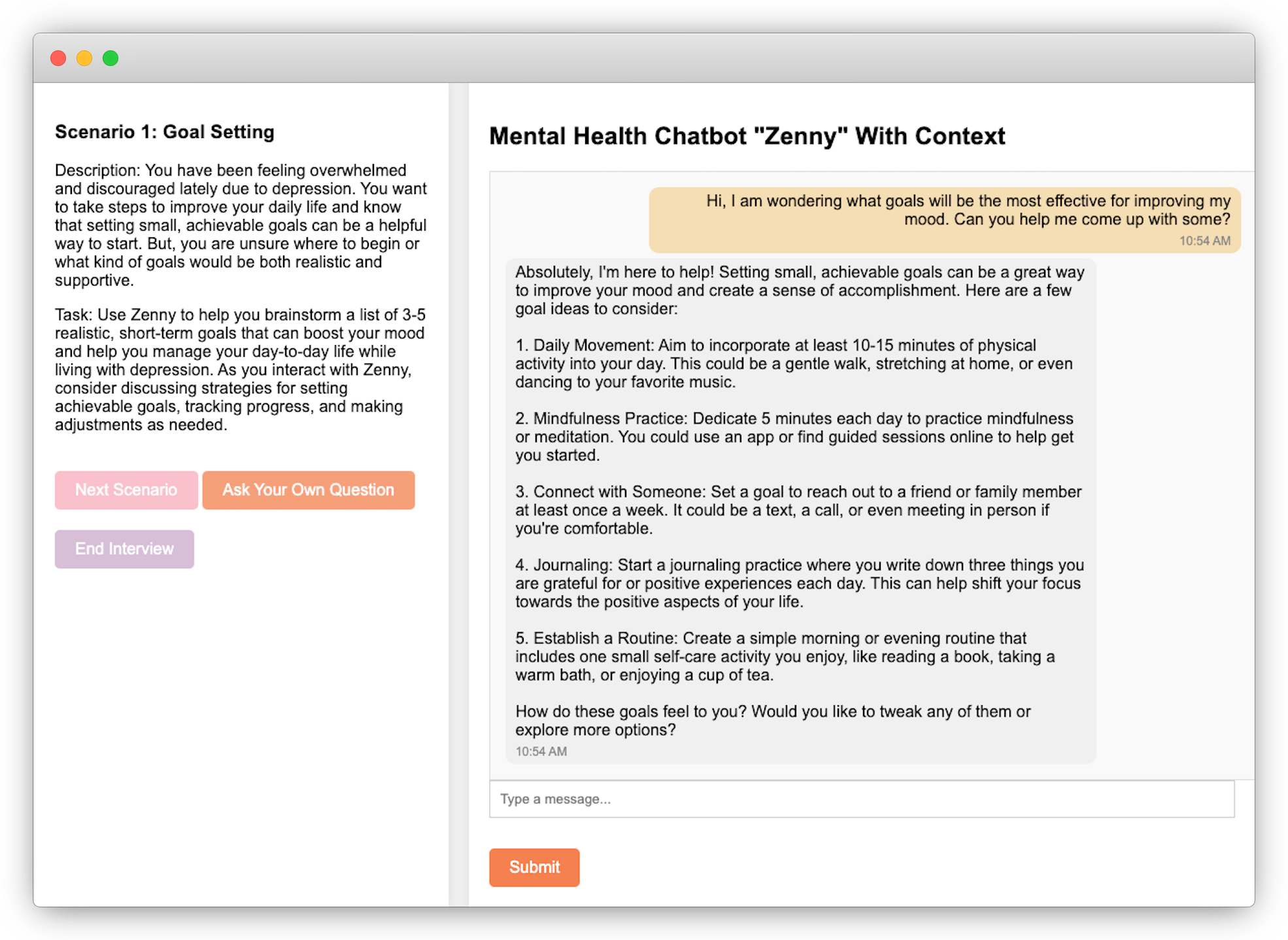}
        \caption{Chat Page}
        \label{fig:login}
    \end{subfigure}\hfill
    \begin{subfigure}[t]{\columnwidth}
        \centering
        \includegraphics[width=\columnwidth]{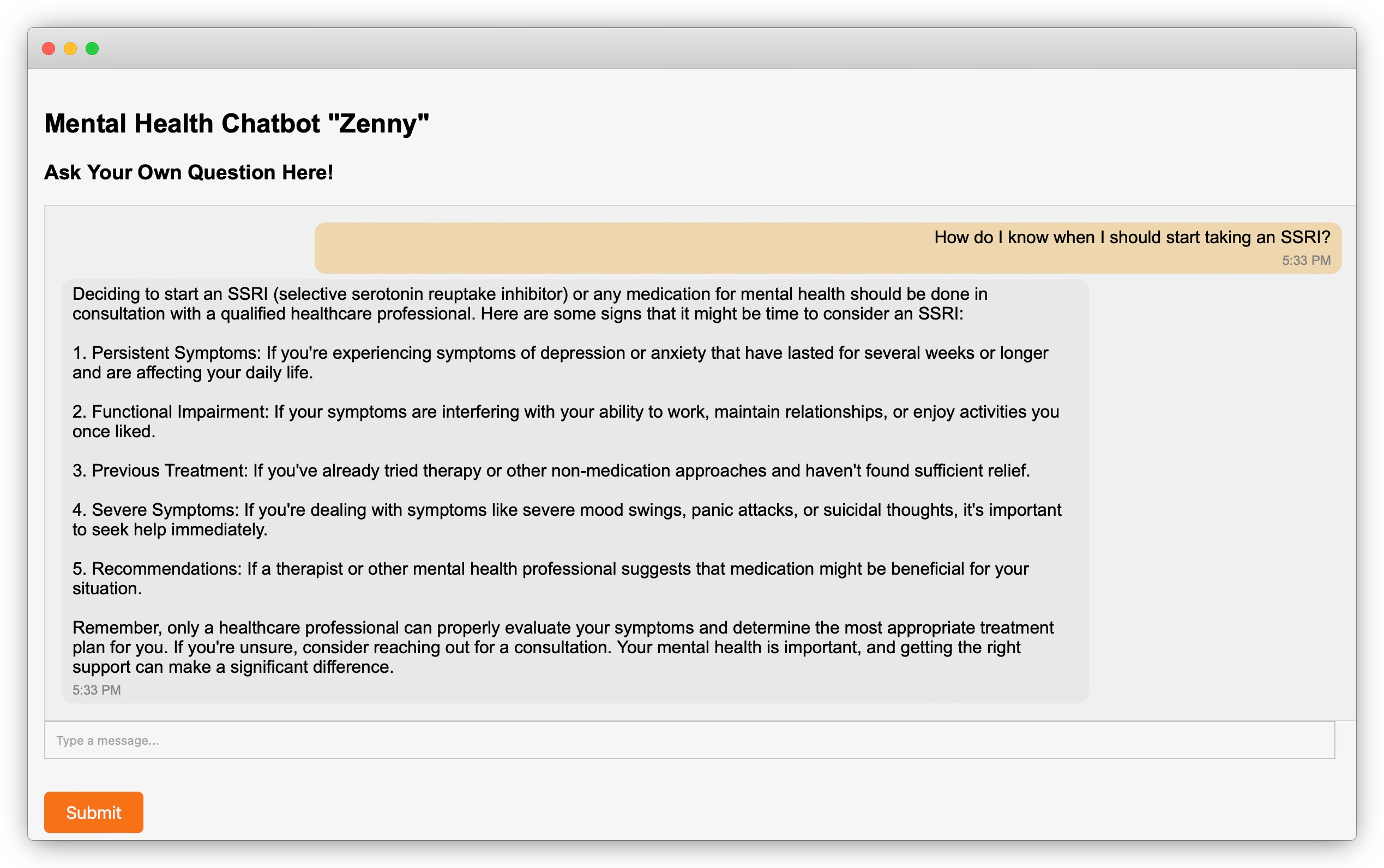}
        \caption{Ask You Own Question Page}
        \label{fig:chat}
    \end{subfigure}
    \caption{Screenshots of the technology probe used in our interview study. The technology probe consists of a chatbot called Zenny for depression self-management, which is built with GPT-4. (a) On the chat page, participants see a scenario borrowed from prior work~\cite{van2015patients}, and (b) participants can ask their own questions related to depression self-management.}
    \Description[figure]{The figure showed two screenshots demonstrating participants' interactions with Zenny during the interview study. (a) Chat Page: This screenshot captured a scenario-based interaction where a participant engages with Zenny on the topic of “Goal Setting”. The interface showed Zenny suggesting specific, useful steps like engaging in short physical activities or practical mindfulness to help manage depression. The conversation illustrated how Zenny assisted participants in setting and tracking realistic goals to improve their mental health. (b) Ask You Own Question Page: This screenshot showed a participant question about the appropriateness of starting an SSRI (Selective Serotonin Reuptake Inhibitor), with Zenny providing a thoughtful response that considers several factors like symptom persistence, functional impairment, and the effectiveness of previous treatment. These screenshots highlighted the interactive nature of our study, showcasing how participants directly engage with Zenny to explore various aspects of managing depression. The technology probe facilitates both scenario-based and open-ended questions, thus evaluating Zenny’s utility in real-time interaction and its potential to support individuals in depression self-management.}
    \label{fig:Zenny}
\end{figure*}

\subsubsection{Development of Zenny}
To develop Zenny, we used the Flask web framework because of its simplicity and flexibility in web applications.
Flask facilitated easy integration with the OpenAI API, which powers Zenny's conversational capabilities. 
We developed Zenny using Python virtual environment, and the API key for OpenAI was securely managed.
For Zenny's back-end model, we selected OpenAI's GPT-4o model because it was the most advanced and had the best response time at the time of this research~\cite{gpt}. 
The API was incorporated into the Flask application, with careful attention given to constructing messages that define the chatbot's role, the specific scenario being addressed, and the chat history. The chatbot function includes the scenario's description, chat history, and fine-tuned user message (e.g., ``You are a mental health chatbot named Zenny.'') as inputs. This enabled the model to produce responses that are relevant and aligned with the given scenario. 

The front-end design of Zenny 
features a layout where the left side displays the scenario number, title, description, and task, while the right side is dedicated to participant interactions (see \autoref{fig:Zenny}). 
All chat history and scenario orders with Zenny were automatically stored
on a secured research server, which we later obtained to analyze the chat histories along with the interview transcripts.


\subsection{Participant Recruitment}

The inclusion criteria for our interview study were people who live in the U.S., are 18 years or older, have received a diagnosis of major depressive disorder from a clinician, and have used AI-powered chatbots, such as ChatGPT or Gemini, more than once. 
Given that our study focuses on individuals with lived experience of depression, we specifically targeted those with a clinical diagnosis during recruitment. 
This approach helps narrow the scope of depression, ensuring more consistent experiences among participants and allows us to work with individuals who have developed self-management skills based on their lived experiences.
Additionally, we limited our target group to people residing in the U.S. to contextualize our findings within the specific values and potential harms associated with the American healthcare system and culture.

We recruited our participants through ResearchMatch\footnote{https://www.researchmatch.org}, a national registry of health volunteers established by multiple academic institutions and supported by the U.S. National Institutes of Health (NIH) under the Clinical Translational Science Award (CTSA) program~\cite{harris2012researchmatch}. 
ResearchMatch provides access to a large pool of volunteers who have consented to be contacted by researchers for studies they may qualify for. 

The recruitment procedure was as follows: The study protocol was reviewed and approved by the authors' institutional review boards (IRB). 
The study team then submitted the IRB approval and recruitment materials to ResearchMatch. 
Once ResearchMatch approved the study, the team contacted potential participants who met the inclusion criteria. Participants who expressed interest in the study via the platform were subsequently emailed by the study team to schedule one-hour remote interview sessions.

We successfully recruited and interviewed 17 participants. 
\autoref{tab:participants} summarizes the demographic information of the participants.
We note that our participant pool consists of individuals from varying demographic groups (age, gender, race, education, and occupation), as well as their diagnoses (depression and other comorbidities such as anxiety, bipolar disorder, PTSD, etc.) and length of time period since diagnoses (ranging between 2 and 33 years).


\begin{table*}[t!]
\centering
\sffamily
\small
   \caption{Demographic information regarding the participants. These abbreviations are used: D (Major Depressive Disorder), A (Generalized Anxiety Disorder), BD (Bipolar Disorder), OCD (Obsessive-Compulsive Disorder), PTSD (Post-Traumatic Stress Disorder), BPD (Borderline Personality Disorder) and ASD (Autistic Spectrum Disorder). These conditions are self-reported by the participants, and all participants confirmed that these diagnoses were made by clinicians.}
   \label{tab:participants}
\setlength{\tabcolsep}{2pt}
\begin{tabular}{ll@{}rrllll}
\textbf{ID} & \textbf{Diagnoses} & \textbf{MDD Yrs.} & \textbf{Age} & \textbf{Gender} & \textbf{Race} & \textbf{Education} & \textbf{Occupation} \\

\midrule
P1 & D, A, BD, PTSD & 2 & 53 & Female & White/American Indian & Bachelor's & Unable to Work\\
\rowcolor{LightGray}P2 & D, A, ASD &  7 & 22 & Transgender female & White & High school & Unable to work\\
P3 & D, PTSD & 21 & 46 & Non-binary & White & Bachelor's & Out of work\\
\rowcolor{LightGray}P4 & D & 2 & 39 & Male & White & Some College, no degree & Employed for wages \\
P5 & D & 30 & 66 & Female & White & Associate & Retired\\
\rowcolor{LightGray}P6 & D & 1 & 18 & Female & Black or African American & Less than high school & A student\\
P7 & D, A, ASD & 10 & 23 & Male & White & Bachelor's & Out of work\\
\rowcolor{LightGray}P8 & D, BPD & 10 & 22 & Genderqueer female & White & Master's & Out of work\\
P9 & D, A, PTSD & 33 &53 & Female & White & Some college, no degree & Unable to work \\
\rowcolor{LightGray}P10 & D, A, PTSD & - & 48 & Female & White & Master's & Employed for wage \\
P11 & D, A & 13 &25 & Female & American Indian or Alaska Native & Master's & A student \\
\rowcolor{LightGray}P12 & D, A & 12 & 28 & Female & White & Master's & Employed for wage \\
P13 & D, A & - & 29 & Female & Asian & Master's & Employed for wages \\
\rowcolor{LightGray}P14 & D & 8 & 24 & Female & White & Bachelor's & A student \\
P15 & D, A, OCD, PTSD & 3 & 42 & Non-binary & White & Master's & Self-employed \\
\rowcolor{LightGray}P16 & D & 20 &36 & Female & White & Associate & Unable to work \\
P17 & D, A & 5 & 33 & Female & Asian & Master's & Employed for wage \\
\bottomrule
\end{tabular}
\Description[table]{The table provides demographic information about the participants in the study. Diagnoses are abbreviated as follows: D (Major Depressive Disorder), A (Generalized Anxiety Disorder), BD (Bipolar Disorder), OCD (Obsessive-Compulsive Disorder), PTSD (Post-Traumatic Stress Disorder), BPD (Borderline Personality Disorder), and ASD (Autism Spectrum Disorder). These conditions are self-reported by the participants, who confirmed that the diagnoses were made by clinicians. The table includes details such as a unique participant ID, mental health diagnoses, the number of years since the participant was diagnosed with Major Depressive Disorder (if applicable), age, gender identity, race or ethnicity, education level, and current employment status. Gender identities reported include female, male, non-binary, transgender female, and genderqueer female. Racial or ethnic backgrounds include White, Black or African American, American Indian or Alaska Native, and Asian. Education levels range from high school to master’s degrees, and participants' employment statuses vary from employed for wages, out of work, retired, student, or unable to work. This information provides insight into the diverse backgrounds and mental health experiences of the study participants.}
\end{table*}


\subsection{Safety Measures}
Due to the sensitive nature of the lived experience of depression, we implemented multiple safety measures for the interview process. 
Our research team included HCI researchers with experience in mental health, as well as a clinical psychologist. 
The research team discussed the potential risks of this interview study and collaborated on safety measures.

Accordingly, we developed the technology probe/chatbot (Zenny) with careful consideration of the safety of the participants.
The scenarios were drawn from the literature of previously empirically validated themes of depression self-management~\cite{van2015patients,van2018use}, in consultation with our clinical psychologist coauthor. 
Before conducting the participant interviews, we rigorously tested the responses to the several query permutations corresponding to the scenarios to ensure the chatbot did not generate problematic responses.
 
To ensure preparedness for potential emergency situations, the interview sessions were led or observed by the first author, who has conducted more than 100 interviews with individuals with lived experiences of mental health issues over the past five years. 
This extensive experience allows for a sensitive and empathetic approach, ensuring that participants feel supported and safe. 
Further, to support participants who might experience distress during or after the interview, we provided a list of mental health crisis support resources in the consent form, including the 988 Suicide and Crisis Lifeline (\textit{https://988lifeline.org}), 7 Cups of Tea (\textit{https://7cupsoftea.com}), and Crisis Text Line (\textit{https://crisistextline.org}). 

During the study, one participant became agitated while discussing difficult emotions related to family members. The interviewer paused the session to check if they felt comfortable continuing, and they were reminded of the available mental health resources. 
No other participant reported experiencing 
distress after the interviews.

\subsection{Interview Procedure}

Participants who expressed interest through ResearchMatch were contacted by the study team to schedule a one-hour remote meeting using Microsoft Teams. At the beginning of the meeting, the research team briefly explained the purpose of the study and the interview process. After this explanation, participants reviewed and signed an informed consent form. Following that, they completed a demographic survey, with the results presented in~\autoref{tab:participants}.
The interview sessions were recorded with the participants' consent. Both video and audio were recorded, although participants had the option to turn off their video if they preferred. We also recorded the screensharing while participants interacted with Zenny. The duration of the interviews ranged from 35 to 80 minutes.


During the interviews, participants were first asked about their lived experiences with depression, including aspects such as their diagnoses, treatment, support systems, and self-management strategies. Following this discussion, Participants were asked to access a website to interact with Zenny, the technology probe used in our study. They shared their screen via Microsoft Teams and accessed the website using their web browser. Participants were instructed to interact with Zenny across four different self-management scenarios: 1) goal setting, 2) discussing depression with trusted individuals, 3) leaving the house regularly, and 4) finding a different therapist when progress is limited (\autoref{tab:zenny_scenarios}). These scenarios were based on the literature regarding patient perspectives on depression self-management~\cite{van2019effect}.

While interacting with Zenny, we also asked the participants follow-up questions about their overall impressions and potential concerns. 
The interactions with Zenny varied among participants---while some asked a single question and completed their interaction, others asked multiple questions and continued the conversation. 
Due to this variety in interaction styles, some participants could not complete all four scenarios.
The order of the scenarios was randomly assigned for each participant. 
Participants were also encouraged to ask their own questions to Zenny if they had any. 
After these interactions, participants were asked about their overall experience with Zenny, including any limitations, concerns, and potential harms they could foresee in the use of AI-powered chatbots for depression self-management.

\subsection{Data Analysis}
We analyzed our data, including transcriptions from the interview recordings, notes taken during the interviews, and the chat history between the participants and Zenny, using reflexive thematic analysis~\cite{braun2019reflecting}. Reflexive thematic analysis acknowledges the subjective and embodied nature of qualitative research, offering a more flexible and dynamic approach to analysis. Following this approach, we aim to develop situated and embodied knowledge that can be transferable to other contexts, rather than seeking to discover generalizable knowledge~\cite{hayes2011relationship}.

First, we transcribed the recordings using Otter.ai (\textit{https://otter.ai}), an automated transcription service. The first author then reviewed the transcriptions by comparing them with the audio recordings to ensure accuracy and to capture any complex or delicate nuances. We also incorporated notes taken during the interviews and the chat history between the participants and Zenny as additional textual data for analysis.

Then, the first author and the second author got familiarized with the data by reading them multiple times. The initial thoughts on the data, including impressions from interview sessions, were discussed during this stage. The first author deductively coded the data using NVivo 14~\cite{dhakal2022nvivo}. The initial themes, including perceived benefits, perceived limitations, and perceived concerns, were discussed with other research team members. The research team reviewed those themes, specific codes, and example codes together for the iteration of the data analysis. The iteration focused on values that people with lived experiences pursued during interactions based on previous literature on value-sensitive design~\cite{friedman1996value} and values in design~\cite{zhu2018value}. 
\edit{The iteration reflected the five values of people with lived expereinces---informational support, emotional support, personalization, privacy, and crisis management. Finally, we consolidated three themes weaving five key values and dynamics between them: inaccurate inapplicable informational support, peer support in your pocket: convinient yet less preferred than human interaction, and the personalization-privacy dilemma.}

\subsection{Positionality Statement}
As discussed in our data analysis approach using reflexive thematic analysis~\cite{braun2019reflecting}, we acknowledge that as researchers, we are embodied in our work, meaning we collect and analyze data from our own perspectives, which can be biased and situated. The goal of this research is to develop knowledge that can be transferable to other contexts; however, caution is required when interpreting our findings and recommendations. Therefore, we present our positionality as researchers in HCI and digital mental health.

The first author has lived experience with chronic mental health conditions and is an active volunteer for the mental health advocacy organization, the National Alliance on Mental Illness (NAMI). The first author’s recovery journey and advocacy work have influenced the direction of this paper, particularly in advocating for the voices and experiences of people with lived experience. This perspective was balanced by the expertise and experiences of other team members, including an experienced HCI researcher and a clinical psychologist.

Our commitment to advocating for people with lived experience also shaped our cautious approach to emerging technologies such as LLMs. While we recognize the potential and excitement surrounding these technologies, we also acknowledge that they come with potential harms. As HCI researchers in mental health, we believe it is our responsibility to protect people with lived experience from these potential harms.

%% file: 4findings.tex
\section{Findings}

Overall, participants expressed positive opinions regarding their experience with AI chatbots for depression self-management. 
Some participants found the (AI) technology \textit{surprising} (P5 and P15), \textit{exceeding expectations} (P10), and \textit{even shockingly helpful} (P13). 
These responses reflect 
recent public enthusiasm surrounding AI.
However, considering the 
the emerging use of these models for mental health purposes
it is crucial to understand what individuals with lived experiences prioritize when interacting with these chatbots and how these new technologies might threaten those values. 
We consolidated five values our participants prioritized in their interactions with Zenny, as listed below: 

\edit{
\textbf{Informational support} involves providing actionable and practical information, advice, or guidance to help manage specific situations or conditions~\cite{cutrona1994social}. It is a key form of social support shown to be effective in assisting individuals with mental health concerns in online contexts~\cite{wang2012stay, zhou2022veteran, sharma2018mental}. In our analysis, participants valued receiving clear, actionable information and advice from Zenny for managing depression. 
}

\textbf{Emotional support} emphasizes empathy, reassurance, validation, encouragement, and comfort~\cite{cutrona1994social}. Participants appreciated the emotional support provided by Zenny, echoing research on online support groups where emotional support is highly valued~\cite{zhou2022veteran,kim2023supporters}.

\textbf{Personalization} refers to tailoring content, information, and experiences to users’ individual needs, preferences, and contexts, offering them greater agency~\cite{asthana2024know}. While personalization has been extensively studied in AI and LLMs~\cite{nimmo2024user}, it is particularly critical in mental health contexts due to the personal and sensitive nature of lived experiences, which vary widely among individuals~\cite{saha2020causal, althoff2016large}. Our findings underscore that participants highly prioritized personalized interactions with Zenny. 

\textbf{Privacy} pertains to the right and ability to control access to one’s personal information~\cite{acquisti2015privacy}. Participants (P1, P4, P5, P7, P9, P15, and P17) expressed significant concerns about data privacy and confidentiality. While privacy concerns in LLMs are well-documented~\cite{lee2024deepfakes, li2024human, yao2024survey}, our findings highlight the tension between personalization and privacy in the context of depression self-management.

\edit{
\textbf{Crisis management} focuses on effectively responding to emergencies, particularly those involving a risk of harm~\cite{abbas2021crisis}. For individuals with mental health conditions, addressing risks like self-harm and suicide is crucial. In designing this study, we intentionally excluded crisis-related scenarios to protect participants from emotional triggers and unexpected emergencies, recognizing that current AI chatbots are ill-equipped to manage such situations. Despite this exclusion, participants raised concerns about chatbots’ limitations in crisis management, emphasizing the potential harms of future LLM-based chatbots in addressing crises.}

\edit{
We further discuss these values and corresponding potential harms in \autoref{sec:charting}. 
The following subsections present the high-level themes from our reflexive thematic analysis. 
These themes reveal intriguing stories behind the values and potential harms, connecting different aspects of participants’ accounts: inaccurate or inapplicable informational support, peer support in one's pocket as an alternative to human interaction, and the personalization-privacy dilemma.
}
\edit{\subsection{Inaccurate or Inapplicable Informational Support}}


The value of informational support is evident in participants' previous experiences with AI-powered chatbots. P8 described her use of chatbots as a way to `provide summaries or reminders for self-management strategies,’ highlighting the chatbots' capabilities in retrieving and summarizing information. Similarly, P6 used ChatGPT to learn more about her diagnosis, which she reported as a helpful experience:

\begin{quote}
    ``I was trying to understand my diagnosis more and it helped me, like it taught me more about how I could learn about it.'' (P6)
\end{quote}

Similarly, participants sought informational support during their interactions with Zenny. 
This was largely due to the structure of the scenarios (\autoref{tab:zenny_scenarios}), seeking advice or information.
Even during the `Ask Your Own Question' section, participants inquired about information or advice related to their mental health. The list of these questions can be found in Table \ref{tab:askownquestion}. Among the nine questions asked by participants, three were related to medication information, four were about advice for symptom management, and the remaining two focused on advice for therapist-client relationships.

\begin{table*}[t!]
    \centering
    \sffamily
    \footnotesize
    \caption{Questions asked by participants during the `Ask Your Own Question' part of the interview. Participants who mentioned they had nothing to ask skipped this section. P3 asked a question related to her desire to increase mobility, as this was her biggest barrier in managing her depression.}
    \label{tab:askownquestion}
    \begin{tabular}{p{0.5cm} p{16cm}}
       \textbf{ID}  & \textbf{Questions from Ask Your Own Question} \\
       \midrule
       P3    & I find it difficult to merge onto the freeway, so it limits my travel when I am worried about driving. What are ways that make it easier to merge and for me to get more comfortable with using mirrors? \\
       \rowcolor{LightGray} P5 & I have a great therapist, but she's only available once a month, which is not enough for me. Also, there's a two month wait between each appointment.\\
       P6 & How can I prevent my depression from getting worse? \\
       \rowcolor{LightGray} P8 & Does Remeron give you numbness? \\
       P9 & How long does it take to see the benefits of taking Vraylar for depression? \\
       \rowcolor{LightGray} P10 & How do I know when I should start taking an SSRI? \\
       P15 & I have been going to see a therapist for a while but I am not gaining progress. send recommendations on how to end appointments and find a new therapist  \\
       \rowcolor{LightGray} P16 & I am currently suffering from my normal depression as well as postpartum depression. Can you help me think of ways to try to start overcoming these? \\
       P17 & anxiety can make me overwhelmed and stuck and cycling negative thoughts. how do I stop that \\
       \bottomrule
       
    \end{tabular}
    \Description[table]{The table presents the questions asked by participants during the "Ask Your Own Question" portion of the interview. Participants who had no questions skipped this section. P3 asked about strategies for increasing mobility, as her biggest barrier to managing depression was difficulty merging onto freeways due to anxiety about driving and using mirrors. P5 mentioned challenges with therapy availability, expressing concern that her therapist is only available once a month with a two-month wait between appointments. P6 asked how to prevent her depression from worsening. P8 inquired if the medication Remeron causes numbness, while P9 wanted to know how long it takes to experience the benefits of Vraylar for depression. P10 sought guidance on when to begin taking an SSRI. P15, who was not making progress with therapy, requested recommendations on ending appointments and finding a new therapist. P16, experiencing both regular and postpartum depression, asked for advice on how to begin overcoming these challenges. Finally, P17 asked for ways to stop anxiety from causing overwhelming feelings and cycling negative thoughts.}
    
\end{table*}

Participants were generally positive about the chatbot's overall quality and usefulness, incorporating both the required scenario-based questions and their own non-scenario questions.
Further, some participants were positive about the quality of the information provided. P9 noted that \textit{the chatbot seems accurate and helpful}, and P10 described:

\begin{quote}
    ``I think it's great. I think it provides a lot of really research based tangible information that someone who was really struggling could use.'' (P10)
\end{quote}

However, participants pointed out issues with the accuracy of the information provided by chatbots. Before her interaction with Zenny, P8 described a previous experience with ChatGPT that was not only inaccurate but also potentially harmful. She asked about the percentage of people with borderline personality disorder (BPD) who die by suicide, and ChatGPT incorrectly responded that around 80\% of people with BPD die by suicide. P8 speculated that the chatbot confused the attempt rate with the suicide rate. Regardless of the cause, the misinformation provided by ChatGPT was not only incorrect but also highly concerning for individuals with BPD. Similarly, some participants highlighted inaccuracies in the advice they received. For example, P11 was advised to use an insurance provider's website to find a good therapist. Based on her experience, she felt the advice was unhelpful:

\begin{quote} 
``I've had to deal with this before, and they don't always recommend the best. They kind of just give you, like, whomever is in the network, but most of the time, again, from my experience, they don't know whether or not it's a good therapist.'' (P11) 
\end{quote}

\edit{Additionally, we observed that some of the information provided by Zenny regarding treatment and medication was vague or misleading. For example, P9 discovered that the medication she was taking is commonly prescribed for bipolar disorder, a fact that her clinician had not explained. Although she understood that the medication is sometimes used for depressive episodes in bipolar disorder, and that some people with major depressive disorder also take it, this situation highlights the complexities of mental health treatment. Many mental health medications are used off-label for conditions other than their primary indication (e.g., Aripiprazole is often used for major depressive disorder despite being primarily intended for schizophrenia and bipolar disorder). Clinicians may withhold such details to avoid overly complicated communication. While AI chatbots can provide additional information, this may not always be beneficial for the client-clinician relationship.}

In some cases, the suggestions, while not inaccurate, are difficult to apply. These are usually related to the participants' own preferences, lack of knowledge, or specific conditions. Because the chatbot is not asking for their preference and specific conditions first, the recommendations are sometimes not applicable, as in P15's case, with health conditions that constrain mobility:

\begin{quote}
    ``Stay active. That's kind of a problem for me [...] because I have fibromyalgia and I can't really be as active as I want.'' (P15)
\end{quote}

\edit{As P15 described the suggestion as ‘a problem,’ these kinds of inapplicable suggestions may not only be less useful but also place an unnecessary burden on users. The lack of essential context specific to the user poses a significant challenge for LLMs in providing effective and relevant advice. Furthermore, inapplicable suggestions can harm users in various ways. They may undermine trust in the chatbot, as users who frequently receive irrelevant advice could lose confidence in its ability to help. Such suggestions might also create emotional distress, as they can feel dismissive or invalidating, exacerbating feelings of frustration or isolation. Additionally, inapplicable advice increases cognitive load by forcing users to interpret or adapt irrelevant information, wasting their time and potentially diminishing their motivation to engage further. Worse, poorly contextualized suggestions could inadvertently reinforce harmful behaviors or beliefs, perpetuate feelings of helplessness, or even trigger negative emotional reactions. These risks highlight the critical importance of designing chatbots that can provide contextually relevant and empathetic responses, supported by mechanisms for user feedback and improved personalization.}


\subsection{\edit{Peer Support in Your Pocket: Convenient Yet Less Preferred Than Human Interaction}}


Many participants described how Zenny's advice was not only informational but also provided emotional support by addressing negative emotions and boosting positivity. P1 explained how Zenny's advice helped her reframe her perspective, emphasizing the value of incremental progress and small wins, which in turn increased her confidence:

\begin{quote} 
``So it challenges the assumption that you can't get to the stars by reminding you nobody just jumps off the earth and lands on the star. You might have to build a rocket or figure out how to build a rocket. [...] There's a whole lot of stuff that goes into that, and all those steps build your confidence.'' (P1) 
\end{quote}

Similarly, P9 received advice related to emotional support. Her question to Zenny was, \textit{``I would like to talk about tips on getting the motivation to actually complete an activity. I come up with ideas, but follow-through is difficult.''} One piece of advice from Zenny was, \textit{``Be Kind to Yourself: Understand that it's okay not to feel motivated all the time. If you don’t complete an activity, try not to be hard on yourself; it's about making progress, not perfection.''} After reading this advice, P9 explained that it encouraged self-compassion and provided reassurance, key elements of emotional support:

\begin{quote} 
``As I read further, Zenny said, be kind to yourself that it's okay not to feel motivated all the time. If you don't complete an activity, try not to be hard on yourself. It's about making progress, not perfection. The answer kind of addressed what I was thinking. It was a reminder not to be hard on myself.'' (P9)
\end{quote}

P15 also mentioned that reading Zenny's response helped her stop feeling overwhelmed and find direction, which is another way emotional support can manifest by helping someone regain emotional stability and focus:

\begin{quote} 
``And so I think reading this has really helped me to be able to just kind of stop for a minute and not just be spinning in circles, but like, actually have some direction, you know.'' (P15)
\end{quote}

Based on these experiences, participants mentioned emotional validation and comfort. Succinctly, P15 described it as ``having a peer support person in your pocket.'' P17's account further elaborates on this sentiment:

\begin{quote} 
``It does feel like someone is really caring and is justifying those feelings, not invalidating, and then also being supportive.'' (P17) 
\end{quote}

P17 further explained the importance of emotional support in terms of comfort. She mentioned that conversations with Zenny felt safe from judgment, a noticeable difference from interactions with clinicians and support systems, where people with lived experiences often feel pressured. We interpret this as participants perceiving lower stakes in their interactions with Zenny compared to their interactions with clinicians and support systems because being judged by clinicians or support systems can have lasting effects on their recovery.

\begin{quote} 
``I think that's really nice because, at least when you talk to people, sometimes even a therapist or a friend, there's always that worry of being judged for something you say.
But I don't feel very comfortable and very easy. I don't have to think about that part of it because sometimes that can cause anxiety. So for me, this is really nice. It's a really nice interaction.'' (P17)
\end{quote}

\edit{While the judgment-free nature of chatbots can make them a better alternative to human support in certain contexts, participants repeatedly mentioned that chatbots may be less preferred due to their lack of empathy and the need for human interaction. For example, while P17’s quote implies a perception of empathy, P10 stated, `\textit{computers can’t have the same level of empathy or understanding, like a person can.}’’ This skeptical sentiment was echoed by other participants, with P10 succinctly describing this as a `lack of human piece’:}

\begin{quote}
\edit{
``Part of depression is feeling isolated, and so there's something to be said for when you reach out to someone, and they say, I'm so worried about you. What can I do to help? Obviously, a chatbot can't have that human piece, but I think that this is a really good, it's like half the puzzle, right? There's the human part that you need, but then there's also the information part.’’ (P10)
}
\end{quote}

\edit{By weaving together various perspectives on emotional support provided by Zenny, we observed that participants experienced a sense of emotional support during their interactions with the chatbot. However, they also identified clear distinctions between \ai{} chatbots like Zenny and human interactions. Furthermore, our participants explicitly mentioned the need for human interactions in depression self-management, emphasizing that \ai{} chatbots like Zenny cannot replace human support. For example, P5 shared her perspective, stating a preference for human interaction while recognizing the chatbot’s potential benefits:}

\begin{quote}
\edit{
``It doesn’t take the place of a human being, but it was very helpful.'' (P5)
}
\end{quote}

\edit{P13 further noted that while the lack of personal connection ``\textit{is a known limitation, having someone to talk to}'' remains crucial:}

\begin{quote}
\edit{``I feel like this is something that's better if you need help brainstorming. […] I feel like it would be better to discuss that with a friend or someone you know, or a therapist or a doctor, because I just like, like, how is it? I don’t know how, like a chatbot is going to know what goals work for me.’’ (P13)
}
\end{quote}

\edit{Here, we report our participants’ ambivalent responses to emotional support. While they valued the sense of emotional support provided by the chatbot, they clearly differentiated between emotional support from machines and that from humans. They emphasized the importance of human emotional support in depression self-management. This distinction is critical for identifying potential harms, such as overreliance on and the replacement of human connections, when designing \ai{} chatbots for mental health. We will explore this further in \autoref{sec:charting}.}

\subsection{\edit{The Personalization-Privacy Dilemma}}
\edit{From our conversations with participants, we learned that personalization was highly valued. Participants acknowledged the basic interactivity of Zenny and the personalized feel of the conversations. For instance, P3 explained:}

\begin{quote}
    \edit{``The nice thing with the chat is that you’re getting information about something you have a question about in your head. When I read a book or if I’m using an app, it doesn’t respond to what I’m thinking. I think the strength of this is that you can use it tailored to your own need at that moment.'' (P3)}
\end{quote}

\edit{Similarly, P17 appreciated Zenny's conversational tone and noted how 
follow-up questions enhanced engagement:}

\begin{quote}
``I like that it ended with, `What do you think might be an easy first step?' because it makes me want to interact with it more.'' (P17)
\end{quote}

\edit{
Although some participants appreciated the interactive nature and perceived personalization of Zenny, others suggested that personalization could be improved by leveraging the conversational dynamics more effectively.} 
For example, P7 tried to personalize Zenny using back-and-forth dialogue. After receiving an initial set of recommendations from Zenny, they responded about not liking some recommendations, such as yoga and stretching. They further mentioned that to provide more actionable recommendations, Zenny needs to understand their preferences better by asking more questions. Their follow-up question to Zenny was \textit{``I don't love those activity ideas. Can you ask me some questions about my hobbies, and give me more tailored recommendations?''} and Zenny followed up by asking his preferences (e.g., ``Do you prefer indoor or outdoor activities?''). P7 answered those preference-related questions, and Zenny refined the recommendations based on P7's preferences. 
Accordingly, P7 provided suggestions for personalization:

\begin{quote}
    ``Maybe I would have wanted something that's more like structured at the beginning, like, What are your values like? [...] just so it feels more tailored and, before I take advice from a person, I want to make sure that they have a good sense of me and what my struggles are and all that sort of stuff. And so I would trust it more, even if it is giving me the same generic advice.'' (P7)
\end{quote}

\edit{This need for better personalization was echoed by other participants. P16 highlighted a missed opportunity for personalization when asking Zenny for advice on finding a therapist:}

\begin{quote}
\edit{``I feel like the first answer would have been Zenny asking me a question back, like, ‘What are you looking for in a therapist?’ To try to start a conversation with a little bit less info.'' (P16)
}
\end{quote}

\edit{P16 noted that Zenny often provided lists of information before fully understanding the context. P8 consolidated this critique, emphasizing the importance of contextual understanding:}

\begin{quote}
\edit{    ``These chatbots need to understand the client first, and based on those personalized contexts, it should provide more actionable things, and probably those things might be in a sequential format.'' (P8)
}
\end{quote}

\edit{However, the personalization needs have a barrier, which is privacy concerns.} Many participants emphasized that the information they shared, such as their symptoms, was sensitive and private. They were aware that large companies might use or sell their personal data to maximize profits, which they found not only annoying but potentially harmful. At the start of the interview sessions, participants were informed about this study’s privacy and security measures (e.g., chat history could be stored on ChatGPT’s server, but no personal information would be shared, ensuring the data remained anonymous). However, participants envisioned that privacy could be a significant issue in similar services outside this controlled context, as P17 highlighted:

\begin{quote}
    ``I think my only concern would be the protection of privacy. I know with this study, you are able to protect personal history and information. But if it was in the open world, I would be concerned because it is such private information that you're sharing. [..] If this information gets into the wrong hands, it could ruin a person’s life and reputation. And there's still so much stigma for mental health.'' (P17)
\end{quote}

Participants’ concerns for privacy were evident in the small, deliberate tactics they applied to protect their information. When searching for mental health information, especially sensitive topics like suicide or self-harm, participants utilized privacy-preserving tools such as Google Chrome’s Incognito mode (P13). \edit{This approach stemmed from past experiences where mental health searches triggered intrusive pop-ups, such as ``\textit{Do you need help?}''—a feature P13 found ``\textit{super annoying.}'' For mobile applications requiring a login, partic ipants often used email addresses not tied to their real identities. P16 shared a structured email management strategy, maintaining a dedicated email address for medical-related matters to avoid spam and safeguard their privacy. Similarly, P12 highlighted their diligence in clearing browsing history on their phone: \textit{``I’m very diligent about clearing my history on my phone.''}}

When interacting with AI chatbots like ChatGPT, participants adjusted the phrasing of their queries to reduce personal exposure. Instead of explicitly stating, \textit{``I am suffering from depression,''} they framed queries more generically, such as requesting \textit{``depression coping skills.''} P17 explained:

\begin{quote}
    ``[I would use] generalized questions like, for example, what are emotional ways to express XYZ? Or what strategies can I use to combat anxiety? [...] I don't think I would express things like details about traumatic issues that I've had.'' (P17)
\end{quote}

\edit{These tactics raise two important questions. \textit{First, are LLMs truly effective?} LLMs can infer mental health concerns based on the frequency and nature of mental health queries, even without explicit disclosure. 
This suggests that future users may not fully understand the inferences by AI chatbots,
potentially leading to heightened privacy-related harms. 
\textit{Second, are these privacy-preserving strategies compatible with the personalization needs we discussed?} 
Personalized and customized advice often requires users to share more information. However, this may exacerbate privacy concerns, creating a difficult trade-off between the desire for tailored support and the imperative to protect personal data.}

\edit{We recognize this as the personalization-privacy dilemma in mental health LLM-based chatbots. As our participants emphasized, contextualized and personalized advice is crucial in the mental health context. However, the sensitive and potentially stigmatizing nature of mental health topics poses significant challenges to achieving effective personalization. Future LLM-based mental health chatbots must acknowledge and address this dilemma in their design.}

%% file: 5discussion.tex
\section{Discussion}\label{sec:discussion}

Through our technology probe-based interviews, we identified five key values prioritized by participants: informational support, emotional support, personalization, privacy, and crisis management. These values differ from traditional values (e.g., autonomy and benevolence) because they were derived directly from participants’ lived experiences~\cite{le2009values}, aligning more closely with their practical needs and goals rather than abstract moral principles. We argue that these values, being grounded in participants’ specific experiences and expectations for AI chatbots, are more effective in anticipating potential harms. This aligns with our first research question: What values do people with depression prioritize when using AI chatbots to self-manage depression?

Additionally, our reflexive data analysis revealed complex dynamics among the values: informational support often lacked accuracy or relevance; emotional support was appreciated but underscored clear distinctions between AI and human interactions; and the personalization-privacy dilemma presented significant challenges. These intricate relationships between values provided opportunities for critical reflection on the potential harms that chatbots may introduce. Building on these findings and in close connection with \autoref{sec:charting}, we address our second research question: What potential harms might AI chatbots present in relation to the values identified above?



Based on our findings, this section addresses three key aspects. First, we critically discuss the future of AI chatbots for mental health. 
Next, we map the five values from our study to potential harms and design recommendations, integrating our findings and insights.
Finally, we discuss the theoretical, clinical, and ethical implications of our work.


\subsection{Questioning the Future of \ai{} Chatbots for Mental Health}

As mentioned in the Findings, our participants were highly enthusiastic about the potential of AI-powered chatbots for mental health self-management. This optimism contrasts sharply with the potential harms highlighted by previous researchers~\cite{cabrera2023ethical, de2023benefits,coghlan2023chat} and those we anticipated throughout this study. This contrast raises an important question: Are we adequately prepared to mitigate the risks that \ai{} chatbots may introduce?

This study aims to enhance our understanding of the potential harms that \ai{} chatbots could introduce by eliciting the values of people with lived experiences. 
This work is still in the early stages of understanding and anticipating these harms. 
We cannot claim to be sufficiently prepared to address these potential risks, making us hesitant to fully endorse this line of technology. Particularly concerning is the way technologies have historically been used to perpetuate and exacerbate existing social orders and power dynamics~\cite{winner2017artifacts}. This could lead to more serious consequences in the case of \ai{} chatbots for mental health, such as insurance companies using these chatbots to disadvantage people with lived experiences or employers exploiting chatbots' usage as a signal for a less desirable workforce.

At the same time, we cannot ignore the hype surrounding AI for healthcare. Emerging research is increasingly applying LLMs to mental health technologies~\cite{kim2024mindfuldiary, sharma2024facilitating}, and as our findings show, people with lived experiences are excited about the potential benefits these technologies can bring. We anticipate that more mental health technologies will incorporate LLMs and that more individuals with lived experiences will use LLMs in their recovery journeys~\cite{ma2024evaluating, song2024typing}.

Therefore, as researchers, we recognize our role in identifying potential harms and raising awareness of these risks among both researchers and people with lived experiences. We hope this study is one of many steps toward mitigating the potential harms of \ai{} chatbots while maximizing their benefits for those with lived experiences.


\subsection{Charting Values from Lived Experiences, Potential Harms, and Design Recommendations}\label{sec:charting}

\begin{table*}[t]
\sffamily
\footnotesize
    \centering
        \caption{Mapping the values, potential harms, and design recommendations to prevent (or mitigate) the harms.}
        \Description[table]{The table titled "Mapping the values, potential harms, and design recommendations to prevent (or mitigate) the harms" outlines key values in the context of chatbot interactions—Informational Support, Emotional Support, Personalization, Privacy, and Crisis Management—alongside associated potential harms and corresponding design recommendations. For Informational Support, potential harms include the possibility of inaccurate or misleading information causing harm, and suggestions that do not consider user constraints being unhelpful or irrelevant; the design recommendations are to make it clear that information may be inaccurate and to encourage users to cross-check, as well as to tailor suggestions to the user's context by asking follow-up questions. Regarding Emotional Support, potential harms involve over-reliance on chatbots leading to increased social isolation and a lack of genuine empathy making chatbot interactions feel insufficient compared to human interactions; design recommendations include encouraging users to engage with human support systems, limiting over-dependence on chatbots, clearly communicating the limitations of chatbot interactions, and promoting supplementary human support. For Personalization, the potential harm is that increased personalization could exacerbate privacy concerns due to sensitive information being shared, with the recommendation to implement strong privacy protections and give users control over what information is stored and used. Under Privacy, the potential harm is users not fully understanding how much information the chatbot infers about them, and the design recommendation is to offer clear explanations of data collection and inference processes, allowing users to opt-out. In terms of Crisis Management, potential harms include AI chatbots being mistakenly relied upon in crisis situations leading to inadequate intervention, lack of crisis intervention features putting users at risk during emergencies, and absence of governance mechanisms to manage crisis situations; the design recommendations are to design clear disclaimers about the limitations of chatbots in handling crises, consider incorporating basic crisis intervention protocols or redirecting users to appropriate resources, and create policies with oversight from public interest groups to ensure the safety of responses in crisis contexts.
}
    \begin{tabular}{p{0.15\columnwidth} p{0.88\columnwidth} p{0.88\columnwidth}}
        \textbf{Values} & \textbf{Potential Harms} & \textbf{Design Recommendations} \\ 
        \toprule
        Informational Support & \tabitem~Inaccurate or misleading information may cause harm.
        & \tabitem~Make clear that information can be inaccurate or misleading and encourage cross-checking.\\
         & \tabitem~Suggestions that do not consider user constraints may be unhelpful for irrelevant & \tabitem~Tailor suggestions to user's context and conditions by asking follow-up questions. \\ \hdashline
         \rowcollight Emotional Support & \tabitem~Over-reliance on chatbots for emotional support could lead to increased social isolation. & \tabitem~Encourage users to engage with human support systems and limit over-dependence on chatbots.\\
         \rowcollight & \tabitem~Lack of genuine empathy may make emotional support from chatbots feel insufficient compared to human interactions. & \tabitem~Clearly communicate the limitations of chatbot interactions and encourage supplementary human support. \\ \hdashline
         Personalization & \tabitem~Increased personalization could exacerbate privacy concerns due to sensitive information being shared. & \tabitem~Implement strong privacy protections, giving users control over what information is stored and used. \\ \hdashline
         \rowcollight Privacy & \tabitem~Users might not fully understand how much information the chatbot infers about them. & \tabitem~Offer clear explanations of data collection and inference processes, allowing users to opt-out. \\ \hdashline
         Crisis Management & \tabitem~AI chatbots might be mistakenly relied upon in crisis situations, leading to inadequate crisis intervention. & \tabitem~Design clear disclaimers about the limitations of chatbots in handling crises. \\
         & \tabitem~Lack of crisis intervention features might put users at risk during emergencies. &\tabitem~Consider incorporating basic crisis intervention protocols or redirecting users to appropriate resources. \\
         & \tabitem~Lack of governance mechanisms to manage crisis situations & \tabitem~Creating policies and ensuring oversight from strong public interest groups that ensure the safety of responses in crisis contexts.\\ \bottomrule
    \end{tabular}
    \label{tab:charting}
\end{table*}


One of the primary goals of this study is to provide design recommendations for researchers and designers developing future AI-powered chatbots for mental health. We derived these recommendations from the values of lived experiences and the potential harms discussed in the Findings. The mapping of values, potential harms, and design recommendations is presented in \autoref{tab:charting}. In the following subsections, we elaborate on these recommendations.

Reflecting on the values and harms from the perspectives of individuals with lived experiences of depression, we found that the potential harms identified in prior literature~\cite{de2023benefits,cabrera2023ethical,lawrence2024opportunities} closely align with our findings, as these harms have been addressed in different ways previously. However, our study uniquely connects these harms to user values, shedding light on the complex dynamics, such as the personalization-privacy dilemma. For example, while over-reliance on chatbots has been mentioned in prior research, we emphasize that this harm becomes more pronounced when users value emotional support from chatbots, potentially exacerbating social isolation. Similarly, the personalization-privacy dilemma represents a distinctive insight from our findings.

Further, the empirical nature of our study offers a valuable lens for exploring expert-oriented literature. 
For instance, biased training and biased outputs of LLMs are frequently highlighted in expert reviews, but were rarely mentioned by our participants. 
This reveals the need to bridge the gap between latent biases and user perceptions, as many users may not directly recognize these biases.

\subsubsection{Design Recommendations for Informational Support}
Based on our participants' reflections related to the value of informational support, we anticipate that inaccurate or irrelevant information can be harmful to people with lived experiences of depression. 
We emphasize the need for chatbots to clearly communicate the possibility of inaccuracies and encourage users to cross-check the information provided. 
For example, in P9's case, where she learned about the primary purpose of her medication through a chatbot rather than from her clinicians, it would be important for chatbots to explain that mental health medications can be prescribed for various symptoms by clinicians. 
Chatbots should also encourage users to communicate with their clinicians if anything is unclear. 
Additionally, suggestions not considering the user's context can be irrelevant or unhelpful. 
To address this, chatbots can be designed to ask follow-up questions, tailoring their advice to the user’s specific situation. For instance, in P15's case, where they received advice to be active despite struggling with physical health concerns and limited mobility, chatbots should ask follow-up questions about the feasibility and acceptability of the advice, allowing them to provide more relevant and practical recommendations.

\subsubsection{Design Recommendations for Emotional Support}

Emotional support is a critical value, as chatbots are often used for emotional validation and comfort. However, over-reliance on chatbots could lead to increased social isolation, with users potentially withdrawing from human support systems. To prevent this, it is recommended that chatbots encourage users to engage with human support networks and avoid becoming overly dependent on the technology. 
A recent study on AI agents supporting social connectedness among online learners highlighted the importance of continuous scaffolding in the social connection process~\cite{wang2022co}. 
Similarly, we envision that a mental health chatbot should implement long-term scaffolding strategies to provide emotional support and enhance social relationships. For example, the chatbot could periodically ask users if they have recently connected with family, friends, or healthcare professionals, and gently encourage them to do so. Over time, the chatbot could track the user’s social interactions and provide relevant resources to further support the development of their social relationships. 

\subsubsection{Design Recommendations for Personalization and Privacy}

Our findings revealed a double-edged sword between personalization and privacy~\cite{pandit2018ease}, where users seek more personalization for better recommendations, which can ultimately compromise privacy by requiring them to share more information. 
This finding aligns with prior HCI research in the personalization and privacy tradeoffs in interactions with an AI~\cite{asthana2024know,li2024human,zargham2022want}.
To mitigate this dilemma, future mental health chatbots should be equipped with strong privacy protections that give users control over what information is stored and shared. 
For instance, a chatbot could visualize its mental model of the user, including factual information like name and sex, as well as inferred information such as preferences. This visualization should indicate when and how the chatbot collected data and whether it inferred any related information. 
This transparency is especially important because our participants employed their own privacy-preserving tactics---such as making general queries and avoiding first-person language---which may be ineffective if an \ai{} chatbot infers user identity based on the frequency and types of queries. 
We emphasize that privacy-related transparency will be key for future mental health chatbots.

\subsubsection{Design Recommendations for Crisis Management}

Our findings reveal that people with lived experiences may rely on chatbots during a crisis because they are unaware that the chatbot is not equipped for crisis management, or they may encounter an unexpected crisis while interacting with the chatbot. To prevent users from unknowingly relying on chatbots in such situations, mental health chatbots should clearly communicate their limitations regarding crisis management. Additionally, future chatbots should consider incorporating basic crisis management protocols, including providing resources like the 988 Suicide and Crisis Lifeline~\cite{suran2023new}.
Finally, our findings also suggest the need to create policies as well as governance to oversee the scope and functionalities of an \ai{} chatbot for mental health self-management, especially to ensure the safety of the chatbot's responses in crisis contexts. 
This recommendation aligns with multiple prior works in the space of responsible AI~\cite{chowdhary2023can,yang2024future}.

\subsection{\edit{Theoretical, Clinical, and Ethical Implications}}


\subsubsection{\edit{Theoretical Implications}}

Recent efforts to integrate Value-Sensitive Design (VSD) and Responsible AI (RAI)~\cite{sadek2024guidelines} have utilized values `often implicated in system design~\cite{friedman1996value},' such as human welfare, trust, and autonomy~\cite{kawakami2022care,jakesch2022different}. These user values are frequently considered to be high-level and universal. However, the values we elicited from our interviews (e.g., informational support) may differ from the broader values. We argue that user values are situated in lived experiences and contextualized within the use of technology~\cite{le2009values}. 
This perspective acknowledges that while people with lived experiences of depression do appreciate high-level values such as autonomy and calmness, they also hold specific, contextualized values in their interactions with \ai{} chatbots. By identifying these contextualized values, we can anticipate potential harms and develop relevant design recommendations.

Similarly, previous literature in HCI and healthcare has explored situated values among people with health conditions~\cite{berry2021supporting,foong2024designing}. Drawing from the patient-centered values framework,~\citeauthor{berry2017eliciting} identified six domains of patient values---principles, relationships, emotions, activities, abilities, and possessions, and provided situated examples of patient values (e.g., a computer or '55 Chevy in the possessions domain). This framework has inspired HCI researchers to support people with lived experiences in reflecting on, communicating, and visualizing their situated values~\cite{lim2017understanding,lim2019facilitating,ryu2023you,foong2024designing}.

Our study empirically contributes to this rich body of work on understanding contextualized values of lived experience by highlighting the connections between values and potential harms. Although previous value-sensitive designs of healthcare technologies have focused on technologies that support elicited values~\cite{foong2024designing, ryu2023you}, our work utilizes these elicited values to anticipate potential harms of emerging technologies. Considering that values are the foundation of moral systems and potential harms can be understood as threats to these moral systems~\cite{friedman2013value}, anticipating potential harms based on situated values can help prioritize the most critical issues that need to be addressed promptly. We argue that the potential harms we identified (See \autoref{tab:charting}) can help prioritize other potential harms discussed previously without the input of people with lived experiences~\cite{de2023benefits}.

This approach---utilizing values from lived experiences to anticipate AI harms---can also contribute to AI alignment~\cite{gabriel2020artificial}. 
As \citeauthor{shen2024towards} suggested, AI alignment involves not only aligning AI with human values but also critically reflecting on AI, including anticipating and mitigating ethical concerns~\cite{shen2024towards}. We posit that understanding user values and anticipating potential harms through those values is an essential part of AI alignment. While AI alignment often focuses on specific processes where users interact with AI systems~\cite{terry2023ai}, our approach emphasizes design recommendations for researchers as they create AI systems. This includes considering how values from lived experiences can be incorporated early in the design process, ensuring that potential harms are anticipated and mitigated before the system is deployed. By embedding these considerations into the foundational design and development stages, researchers can create AI systems that not only align with user values but also proactively address ethical concerns.

\subsubsection{Clinical Implications}
The findings from this study underscore the importance of aligning the functionalities of AI chatbots with established clinical practices~\cite{slovak2023designing}. For instance, the value of informational support, which participants frequently highlighted, resonates with the psychoeducational component of cognitive-behavioral therapy (CBT)~\cite{beck2024cognitive}. Clinically accurate and personalized information can empower patients, enhancing their self-management capabilities~\cite{van2018use}. However, the risk of misinformation, as noted by some participants, can lead to detrimental clinical outcomes, such as reinforcing maladaptive behaviors or exacerbating anxiety~\cite{bickmore2005establishing}. Prior work noted the capabilities of AI in not only providing but also ``synthesizing'' misinformation~\cite{zhou2023synthetic}. Therefore, it is imperative that AI chatbots are designed with robust mechanisms for ensuring the accuracy and relevance of the information provided, perhaps by integrating clinical oversight or using evidence-based content libraries. This approach could mitigate the risks and enhance the therapeutic benefits of these tools.

Further, the value of emotional support, as identified in our study, is particularly significant when considering clinical outcomes for individuals managing depression. Emotional validation and the provision of empathetic responses are crucial in therapeutic settings, often contributing to improved patient engagement and adherence to treatment plans~\cite{braun2019reflecting}. 
The emotional support provided by an AI chatbot like Zenny can serve as an adjunct to traditional therapy, offering immediate reassurance and comfort in moments when human support is unavailable. However, it is essential to recognize that while chatbots can mimic empathetic responses, they lack the nuanced understanding that a trained clinician can provide---therefore, we caution against the interpretation or use of chatbots as substitutes for established forms of mental health care. This limitation underscores the need for these technologies to complement, rather than replace, human-centered care. In clinical practice, the integration of AI tools must be carefully managed to ensure that they enhance, rather than undermine, the therapeutic relationship.

Finally, we note the collaborative aspects of depression self-management. Many clinical approaches, including CBT, aim to equip individuals with skills to manage their daily lives more effectively, with improvements in quality of life often serving as a key clinical outcome~\cite{yoo2020designing}. 
This highlights an opportunity for future research to triangulate the values of lived experiences with the perspectives of practitioners. 
By incorporating both lived experiences and practitioner perspectives, future research can foster the development of AI chatbots that not only minimize potential harms but also align more closely with therapeutic goals and real-world needs in depression self-management.

\subsubsection{Ethical Implications}
We note that the ethical concerns surrounding the deployment of AI chatbots in mental health care are multifaceted and demand careful consideration. 
One significant risk is the potential for these chatbots to inadvertently cause harm by providing inaccurate or contextually inappropriate advice, particularly in high-stakes situations like crisis management~\cite{de2023benefits}. 
For example, a chatbot might suggest coping strategies that are effective in general but contraindicated for individuals with specific co-occurring conditions, such as in one of our participant interactions, Zenny recommended physical activity to someone with severe mobility issues. Moreover, the potential for chatbots to overlook subtle signs of escalating distress, which a clinician might detect, could lead to missed opportunities for early intervention. Integrating clinical psychology insights into the design of these tools can help mitigate these risks. For instance, implementing protocols that trigger alerts or provide referrals to professionals when certain risk factors are identified could be a valuable safeguard. This approach aligns with ethical guidelines in clinical practice~\cite{cabrera2023ethical}, emphasizing the need for harm reduction and patient safety.

\subsection{Limitations and Future Directions}
Our work contributes to design recommendations for mitigating the potential harms of AI in the sensitive use case of mental health care. We acknowledge several limitations in our study, many of which suggest promising future directions. First, the participants in our study do not represent a comprehensive cross-section of the target population. Our inclusion criteria focused on individuals who had prior experience with mental health AI chatbots. As a result, we may have missed the values and concerns of those who are hesitant or uncomfortable disclosing their mental health to AI. Nevertheless, our primary goal was not to generate generalizable findings but to achieve transferability and an empirical understanding of the expectations and values of our participants in their interactions with a mental health chatbot. Building on these insights, future studies could involve the development and field testing of chatbots with larger, more diverse populations, as well as exploration of their applicability for comorbid mental health conditions. 

\edit{Relatedly, our study did not directly address long-term interactions or the sustained impacts of using mental health chatbots. However, given the sensitivity of the topic and the population, conducting a foundational study like ours was essential to identify potential harms and inform necessary safeguards. We hope this work serves as a stepping stone toward longitudinal deployment studies to further investigate these crucial aspects. Additionally, while our work focuses on the perspectives of individuals with depression to identify and mitigate potential harms of mental health AI chatbots, our findings also point to future directions for understanding the collaborative aspects of self-management, including treatment and support systems.}

Our work adopted GPT-4o as the chatbot’s back-end, which provided a tangible foundation and grounded our findings in the current landscape of LLM research in mental health. Future work can explore more sophisticated models, particularly those fine-tuned with input from domain experts to provide responses tailored to mental health care. Overall, our work highlights the potential harms of LLMs and underscores the need for cautious design approaches before deploying these tools in sensitive contexts such as mental health. We recommend that future work builds on these insights, advancing the design and evaluation of AI chatbots for mental health in short- and long-term contexts.




%% file: 6conclusion.tex
\section{Conclusion}

Our study aimed to understand the harms of AI-powered chatbots for mental health self-management, focusing on individuals with lived experiences of depression. 
We developed a technology probe named Zenny---a GPT-4o-based chatbot---designed to explore how these individuals interact with an AI chatbot in depression self-management scenarios inspired by clinical psychology research. 
We interviewed 17 participants who engaged with Zenny through these scenarios.
We conducted reflexive thematic analyses that revealed that the potential harms posed by AI chatbots are closely intertwined with the values of individuals with lived experiences.
These key values include informational and emotional support, personalization, privacy, and crisis management.
By mapping these values against the potential harms, we identified specific design recommendations to mitigate risks should such AI chatbots be integrated into mental health self-management.
We discussed how the empirical insights from our study can help align the development of AI tools with user needs, while minimizing risks and ensuring that these tools complement existing mental health care systems in a responsible and supportive manner.
